\documentclass[aps,twocolumn,showpacs]{revtex4}
\usepackage{graphicx}
\usepackage{float}
\begin{document}
\def\r{{\bf{r}}}
\def\k{{\bf{k}}}
\def\kt{{\tilde{\k}}}
\def\mG{{\hat{G}}}
\def\mg{{\hat{g}}}
\def\mGa{{\hat{\Gamma}}}
\def\mS{{\hat{\Sigma}}}
\def\mT{{\hat{T}}}
\def\bG{{\bar{G}}}
\def\mbG{{\hat{\bar{G}}}}
\def\tk{\tilde{\bf{k}}}
\def\ep{\epsilon}
\def\ep{\epsilon}
\def\de{\delta}
\def\De{\Delta}

\title{Dynamics of Impurity and Valence Bands in Ga$_{1-x}$Mn$_{x}$As \\ 
within the Dynamical Mean Field Approximation}

\author{M.~A. Majidi$^{1}$, J. Moreno$^{1,2}$,M. Jarrell$^{1}$, 
R.~S. Fishman$^{3}$, and K. Aryanpour$^{1,4}$}
\affiliation{$^{1}$Department of Physics,
University of Cincinnati, Cincinnati, Ohio 45221}
\affiliation{$^{2}$Department of Physics, University of North Dakota,
Grand Forks, ND 58202}
\affiliation{$^{3}$Condensed Matter Sciences Division, Oak Ridge National Laboratory, 
Oak Ridge, Tennessee 37831}
\affiliation{$^{4}$Department of Physics, University
of California, Davis, California 95616}
\date{\today}

\begin{abstract}
We calculate the density-of-states and the spectral function
of Ga$_{1-x}$Mn$_{x}$As within  the
dynamical mean-field approximation. Our model includes the competing
effects of the strong spin-orbit coupling on the $J=3/2$
GaAs hole bands and the exchange interaction between the magnetic
ions and the itinerant holes. We study the quasi-particle and
impurity bands in the paramagnetic  and ferromagnetic phases for
different  values of impurity-hole coupling, $J_c$, at the Mn doping
of $x=0.05$.  By analyzing the anisotropic  angular distribution of
the impurity band carriers at $T=0$, we conclude
that the carrier polarization is optimal when the carriers move
along the direction parallel to the average magnetization.
\end{abstract}

\pacs{75.50.Pp,2.70.-c,5.10.-a,71.27.+a}

\maketitle

\par

The combined magnetic and semiconducting 
characteristics of ferromagnetic semiconductors
make them excellent candidates for spintronic applications \cite{Wolf01}.
In particular, GaAs doped with Mn (Ga$_{1-x}$Mn$_{x}$As)
is promising as a spin-carrier injector in spintronic devices\cite{Ohno99} due to its relatively
high magnetic transition temperature \cite{MacDon05} 
and its potential integration within the current semiconductor technology. 
However, the properties of magnetic semiconductors still need to be greatly optimized
since practical uses demand magnetic transitions above room temperature and
carrier polarization of at least 70\% \cite{Wolf01}.

A realistic model which incorporates the relevant bands of the host material is
crucial to guide the experimental efforts in the search for optimal
magnetic semiconductors. In Ga$_{1-x}$Mn$_{x}$As, the Mn ions are in
the Mn$^{2+}$ state with a half-filled  $d$ shell of total spin
$S=5/2$ \cite{Linnarsson97,Okabayashi98}.
Since Mn$^{2+}$ ions primarily replace Ga$^{3+}$, they contribute carrier holes to 
the $p$-like valence band. The strong spin-orbit interaction couples the $l=1$
angular momentum to the electron spin ($s=1/2$), resulting in
a total spin $J=|l+s|=3/2$ for the two upper valence bands and
$J=|l-s|=1/2$ for the split-off band \cite{blakemore}.
Since the $J=3/2$ bands are degenerate at the
$\Gamma$ point, an accurate model should include
at least these two bands. However, a more realistic approach should
incorporate the split-off and conduction 
bands as we discuss later.

Here, we continue our dynamical mean-field approximation (DMFA)
\cite{metzner-vollhardt,muller-hartmann,pruschke,georges} study
of the effects of strong spin-orbit coupling in 
Ga$_{1-x}$Mn$_{x}$As\cite{Aryanpour04}.  While we previously examined the 
influence of the spin-orbit interaction on the ferromagnetic
transition temperature $T_c$ and the carrier polarization
\cite{Aryanpour04}, we now focus on the density-of-states, the
spectral function, and the dispersion of the quasi-particle and
impurity bands.  We also discuss the anisotropy of the spectra in
the ferromagnetic phase and its influence on the transport
properties.

Although the formation of the impurity band has been
captured in previous DMFA studies \cite{chattopadhyay},
their model does not take into account the spin-orbit coupling and
is unable to address the reduced carrier polarization within the impurity band.  
The DMFA describes the impurity band through quantum
self-energy corrections which are not included in other mean-field
theories.  Because this method is non-perturbative, it allows us to
study both the metallic and impurity-band regimes as well as both
small and large couplings. Although the precise role played by the impurity band
in Ga$_{1-x}$Mn$_{x}$As is still controversial, an array of experimental probes,
such as  angle-resolved photoemission \cite{Okabayashi01},
infrared spectroscopy \cite{singley02,singley03,burch05},
spectroscopic ellipsometry \cite{burch04},
scanning tunneling microscopy \cite{Grandidier00,Tsuruoka02},
and photoluminescence techniques \cite{Sapega05}, display features 
characteristic of an impurity band.

We start with the Hamiltonian proposed in
Ref.~\onlinecite{Aryanpour04} and 
\onlinecite{janko}
:
\begin{equation}
H=H_0 - J_c \sum_{R_i} {\bf S}_i \cdot \hat{\bf
J}(R_i)\ .
\label{H}
\end{equation}
The first term incorporates the electronic dispersion and the spin-orbit
coupling of the $J=3/2$ valence holes within the spherical 
approximation\cite{Baldereschi73}. The second term represents the interaction
between the Mn spins and the valence holes \cite{kacman}, with $J_c$
the exchange coupling and $\hat{\bf J}(R_i)$ the total $J=3/2$ spin
density of the holes at the site $i$ of a Mn ion with spin ${\bf S}_i$.  
The relatively large magnitude of the Mn spin  ($S=5/2$) justifies its
classical treatment.

Since the typical hole concentration is
small (around $5\%$), the holes gather
at the J=3/2 bands
around the $\Gamma$ point. This supports the
use of the spherical
approximation \cite{Baldereschi73},
for which the non-interacting
Hamiltonian of pure GaAs is
rotationally invariant. Hence, $H_0$ is
diagonal in a {\it chiral} basis,
$\displaystyle
H_0=\sum_{\k,\gamma}\frac{k^2}{2m_{\gamma}}{\tilde{c}^{\dag}_{\k\gamma}}
\tilde{c}_{\k\gamma}\,$,
where $\tilde{c}^{\dag}_{\k,\gamma}$ creates
a chiral hole with momentum $\k$
parallel to its spin and 
${\bf J}\cdot\hat{\k}=\pm 3/2$ or $\pm 1/2$. The two
band masses
$m_h\approx0.5m$ and $m_l\approx0.07m$
correspond to the heavy and
light bands with $\gamma=\pm 3/2$ and $\pm 1/2$
respectively ($m$ is
the electron mass). Notice that for convenience we use the hole
picture 
so that the valence bands have a minimum instead of a
maximum at zero momentum.  
However, in displaying the results we
reverse the sign back to accommodate the usual convention.

As
discussed previously \cite{Aryanpour04}, the coarse-grained Green
function matrix in the non-chiral fermion basis
is
\begin{equation}
\label{eq:non-chiral-gf}
\hat{G}(i\omega_n)=\frac{1}{N}\sum_{\k}[i\omega_n\hat{I}-\hat{\epsilon}(k)+\mu
\hat{I}-
\hat{\Sigma}(i\omega_n)]^{-1}\,,
\end{equation}
where $N$ is the number of $\k$ points in the first Brillouin zone, $\mu$
is the chemical potential and
$\displaystyle
\hat{\epsilon}(k)=
\hat{R}^{\dag}(\hat{\k})
\frac{k^2}{2m_{\gamma}}\hat{R}(\hat{\k})\,$
is the dispersion in the
spherical approximation. Here, $\hat{R}$ are spin
$3/2$ rotation matrices that relate the fermion operator $c_{\k\gamma}$
to its chiral counterpart $\tilde{c}_{\k\gamma}=R_{\gamma \nu}(\hat{\k})c_{\k \nu}$,
and repeated spin indices are summed. The mean-field function
$\hat{\cal{G}}(i\omega_n)=[\hat{G}^{-1}(i\omega_n)+\hat{\Sigma}(i\omega_n)]^{-1}$
is required to solve the DMFA impurity problem. At a non-magnetic site,
the local Green function equals the mean-field function
$\hat{G}_{non}=\hat{\cal{G}}$, while  the local Green function at a magnetic site is
$\hat{G}_{mg}(i\omega_n)=[\hat{\cal{G}}^{-1}(i\omega_n)+J_{c}
{\bf S}\cdot\hat{\bf J}]^{-1}$.  To obtain this result, we treat disorder
in a fashion similar to the {\it coherent potential approximation} (CPA) \cite{taylor}.

Now $\hat{G}_{mg}(i\omega_n)$ must be
averaged over all possible spin
orientations at the local site and
over all possible impurity 
configurations on the
lattice. The
former is implemented by introducing the angular distribution function
$\displaystyle P({\bf S})=\frac{\exp[-S_{eff}({\bf S})]}{\cal Z}$,
where
${\cal Z}=\int d\Omega_{\bf S}\exp[-S_{eff}({\bf S})]$
and
$S_{eff}({\bf S})$ is the effective action of the system
\cite{furukawa}

\begin{eqnarray}
\label{eq:eff-action}
S_{eff}({\bf
S})=-\sum_{n}\log\det \left[ \hat{\cal{G}}(i\omega_n)
(\hat{\cal{G}}^{-1}(i\omega_n)+J_{c}{\bf S} \cdot\hat{\bf J})\right]
\nonumber\\&&\hspace*{-8.0cm}\times e^{i\omega_n0^+}.
\end{eqnarray}
The extra factor of $\hat{\cal{G}}(i\omega_n)$ in
Eq.(\ref{eq:eff-action}) is introduced to aid in convergence.
If the Mn ions are randomly distributed with probability $x$, then
the configurationally-averaged Green function reads
$\displaystyle
\hat{G}_{avg}(i\omega_n)=\left\langle\hat{G}_{mg}(i\omega_n)\right\rangle
x+\hat{\cal{G}}(i\omega_n)(1-x)$.

When the magnetic order is along the z axis,
$\hat{G}_{mg}$, $\hat{\cal{G}}$ and
$\hat{\Sigma}$ are diagonal matrices and the angular distribution
function depends only on the polar angle of the impurity spin:
$P({\bf S})=P(\theta)$ \cite{Moreno05}.  In fact, 
$\hat{G}_{mg}$ can be written as a
combination of the identity matrix, $\hat{I}$, 
$\hat{J_z}$,
$\hat{J_z^2}$ and $\hat{J_z^3}$, where the coefficients are of order zero,
one, two and three, respectively, of the magnetization:
$\displaystyle 
\hat{G}_{mg}= O(1) \hat{I} + O(M)
\hat{J_z} + O(M^2)\hat{J_z^2}+
O(M^3)\hat{J_z^3}\,$ \cite{Moreno05}.
Making use of these results a new algorithm has been designed which
dramatically reduces the computational time compared to our
earlier work \cite{Aryanpour04}.  
This improvement is due to the fact that  the rotational symmetry of the
model is explicitly broken 
by our choice of a preferential direction for the magnetization.
Also, since the symmetry breaking is already incorporated in this algorithm, 
we no longer need a small magnetic field to break the symmetry along 
a magnetization axis \cite{Aryanpour04}.

To calculate dynamical quantities, we work in the real frequency domain,
where the coarse-grained Green function matrix
is
\begin{equation}
\mG(\Omega)=\frac{1}{N}\sum_{\bf k}[\Omega
\hat I - {\hat \epsilon}({\bf k}) - \mS({\bf \omega})]^{-1}\,
\label{Glocal}
\end{equation}
with $\Omega=\omega + i0^+$.

Generally the Matsubara and real frequency Green functions need to be iterated
simultaneously and averaged over the polar angle distribution
$P(\theta)$. However, if we focus only on the 
paramagnetic phase ($T>T_c$) and the ferromagnetic ground state ($T=0$)
the angular distributions are known and self-consistency in the Matsubara domain
is not  necessary. In the paramagnetic phase, the angular
distribution of the Mn spins is completely 
random so that $P(\theta)=1/\pi$. 
In the $T=0$ ferromagnetic ground state, 
the average impurity magnetization achieves its full value
and $P(\theta)=\delta(\theta)$.

After the coarse-grained Green function is self-consistently calculated, 
the density-of-states is computed as 
\begin{equation}
DOS(\omega)=-\frac{1}{\pi}\ {\rm Im}\
Tr \ \mG(\Omega)\ ,
\label{DOStotal}
\end{equation}
where $Tr$ is the trace. Each of the diagonal elements of 
$\displaystyle -\frac{1}{\pi}$ Im $\mG(\Omega)$ 
is the projection of the density-of-states onto a state
with fixed $J_z$ component, i.e. $J_z=+3/2,+1/2,-1/2$, and $-3/2$.

We are also interested in the spectral function defined as 
\begin{equation}
A({\bf k}, \omega)= -\frac{1}{\pi}\ {\rm Im}\ Tr\ [\Omega{\hat I} - 
{\hat \epsilon}({\bf k}) - \mS(\omega)]^{-1}\ .
\label{Akw}
\end{equation}
The center of the quasi-particle peak in the spectral function represents the renormalized 
quasi-particle energy $\omega_{\mu}({\bf k})$ ($\mu=1,2,3,4$), which can be obtained by solving 
the condition:
\begin{equation}
{\rm Re}\ [\Omega{\hat I} - 
{\hat \epsilon}({\bf k}) - \mS(\omega)]_{diag}=0\ ,
\label{Wk}
\end{equation}
where the subscript $''diag''$ means that we first diagonalize the matrix, then solve the 
equation for each diagonal element.

\begin{figure}
\includegraphics[width=3.2in]{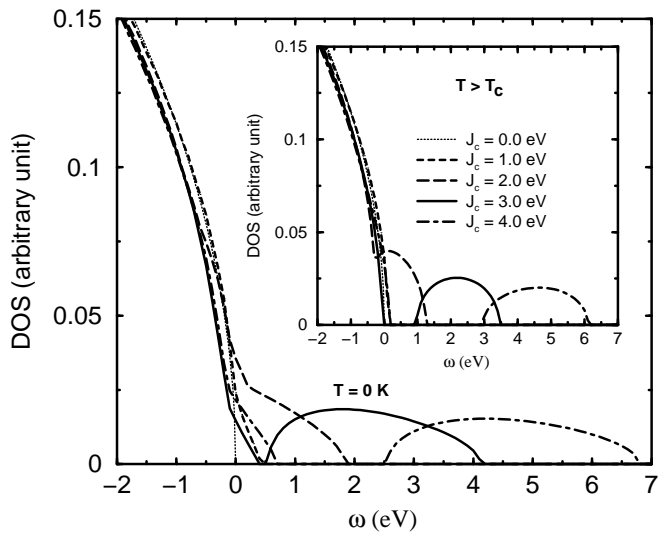}
\caption[a]{Density-of-states for $J_c$=1.0, 2.0, 3.0, and 4.0 $eV$, 
at $T=0$ (main panel) and $T>T_c$ 
(inset). The appearance of an impurity band is evident for $J_c > 2.0 \ eV$. The broader impurity 
band at $T=0$, compared to that at $T>T_c$, indicates that lifetime of the bound-state holes 
decreases as the system becomes magnetically ordered.} 
\label{DOSJc1to4}
\end{figure}

We focus on the doping of $x=0.05$, for which $T_c$ is near the highest 
reported \cite{mbe,Ohno92,Ohno96,VanEsch97,Ohno98,Chiba03,Edmonds04,MacDon05}.  Fig.~\ref{DOSJc1to4} 
shows our results for the density-of-states  for various values of $J_c$ 
in the ferromagnetic phase (main panel) and the paramagnetic phase (inset). As $J_c$ increases,
states with positive energy appear inside the semiconducting gap.
These states correspond to the Zeeman splitting of the hole levels induced by the local
impurity magnetization. For $J_c>2.0\ eV$ an impurity band clearly appears.
For $J_c=3.0\ eV$ and $T=0$, a 
second impurity band has started to form though it has not yet separated from the main band.
We also observe a second impurity band appearing at positive energies 
in the paramagnetic phase when  $J_c>5.0\ eV$ (not shown in the graphs). 
The appearance of two impurity bands is consistent with 
the fact that the model includes two bands with ${\bf J}\cdot\hat{\k}=\pm 3/2$ and $\pm 1/2$.

As expected, the center of the impurity band shifts to higher energy as 
$J_c$ increases.  However, our predictions for 
the energy of the impurity band are too large. 
We believe that this is a consequence of 
excluding the conduction band from our model, since band repulsion
with the conduction band pushes the impurity band to lower energies.
We also notice that the impurity 
band in the ferromagnetic phase is broader than the one in the 
paramagnetic phase, confirming previous results \cite{chattopadhyay,furukawa}. 
This suggests that the lifetime of the ``bound-state'' 
holes is shorter in the ferromagnetic phase because additional
scattering events that exchange holes between 
impurities are required to maintain magnetic order.  Hence, we will refer to
the impurity-band states as {\it quasi-bound} states.

Next, we compare our density-of-states results 
with ARPES data in the paramagnetic phase of Ga$_{1-x}$Mn$_{x}$As, 
$x=0.035$ \cite{Okabayashi01}. 
For this doping Okabayashi {\it et al.} observe an impurity band already well 
separated but not very far apart from the main band. 
A rough estimate for $J_c$ of $2.0-3.0\ eV$  is the most suitable to describe 
the situation observed in the experimental setup. 
In the following  discussion we take $J_c=3.0\ eV$, though this value is an 
overestimate of the exchange coupling, to further explore the physical consequences 
of our model and the behavior of the impurity band.

\begin{figure}
\includegraphics[width=3.2in]{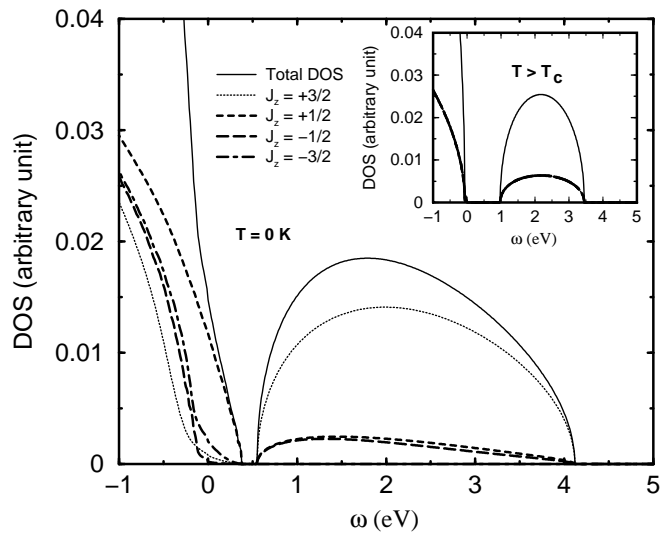}
\caption[a]{Decomposition of the density-of-states in terms of $J_z$ components, for $J_c=3.0 \  eV$ 
at $T=0$ (main panel) and $T>T_c$ (inset). The chiral nature of the holes due to spin-orbit 
coupling mixes up states with different $J_z$ components, making the total DOS only partially 
polarized at $T=0$.} 
\label{DOS.Jc3}
\end{figure}

Fig.~\ref{DOS.Jc3} shows the decomposition of the DOS in terms of its $J_z$ components for $J_c=3.0\ 
eV$. As expected, at $T>T_c$ all four components of $J_z$ contribute equally to the total DOS, 
the electronic system is unpolarized.
At $T=0$, however, we see that the impurity band is not fully 
polarized, as would be expected for the double exchange model.  
In addition to the dominant $J_z=+3/2$ component, components 
with $J_z=+1/2$ and $-1/2$ are also present.  
This is clearly a consequence of the strong spin-orbit coupling, which
mixes the $J_z=+3/2$ state with $J_z=\pm 1/2$ states. 
Previous DMFA studies\cite{chattopadhyay} with coupling to only
one carrier band captured the formation of the impurity band but 
were unable to address the effect of frustration on the carrier polarization.

\begin{figure}
\begin{center}
\includegraphics[width=3.5in]{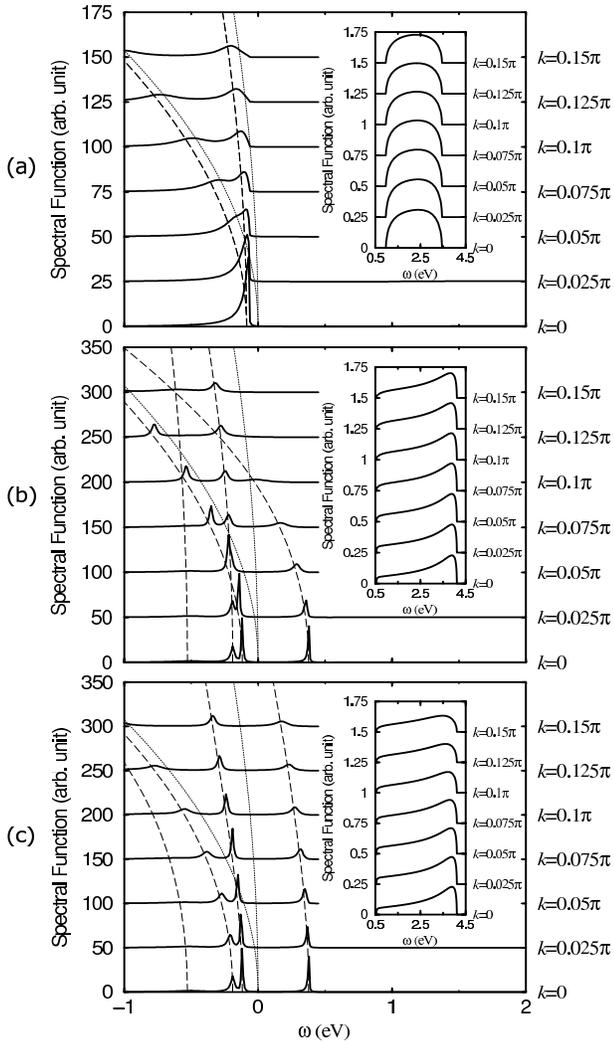}
\caption[a]{Spectral functions near the center of the Brillouin zone for $J_c=3.0\ eV$ 
at (a) $T>T_c$ along any direction, (b) $T=0$ along the direction parallel and (c) perpendicular  
to the magnetization. Insets display a zoom of the 
impurity band region. Each spectral curve corresponds to a different value of the magnitude
of the momentum {\it k}, as indicated at the right of the graphs, and 
the baseline of each spectrum is shifted proportionally to {\it k} for clarity. 
In the main panels the dashed curves represent the 
renormalized valence bands. For comparison, the valence bands of pure GaAs are also displayed 
(dotted curves). The scale in the insets is blown-up 
by a factor of 100 in (a) and 200 in (b) and (c).}
\label{spectragraphs}
\end{center}
\end{figure}

Now we explore the spectral function of our model for 
$J_c=3.0\ eV$. Fig.~\ref{spectragraphs} (a) shows the spectral function in the paramagnetic phase.
Figs.~\ref{spectragraphs}(b) and (c) display the $T=0$ spectrum along the direction parallel and 
perpendicular to the average magnetization, respectively.  Notice 
two main effects on the energy levels:  the valence band quasi-particle states (shown in main panels) 
are renormalized and the impurity band (shown in the blown-up insets) appears. 
Due to the localized nature of the impurity band states, their spectral 
weight extends over a large region in momentum space with typical values of the spectral 
function reduced by two orders of magnitude in comparison with the quasiparticles peaks.

The spectrum within the paramagnetic phase is isotropic, while it is obviously
anisotropic in the ferromagnetic phase. For $J_c=3.0\ eV$ all the 
quasi-particle lines track the peaks, and quasi particles are well defined.
As expected in the paramagnetic phase, the states
corresponding to  ${\bf J}\cdot{\hat{\bf k}}=\pm3/2$ and $\pm1/2$ remain degenerate
for all values of momentum.
In this phase, the self-energy matrix is proportional to the identity \cite{Moreno05},  
preserving 
${\bf J}\cdot{\hat{\bf k}}$ as a good quantum number. The self energy just shifts 
the quasiparticle bands towards negative energies due to
the band repulsion between the quasiparticle and the emerging impurity band. Also notice 
that the heavy and light quasiparticle bands are still 
degenerate at the $\Gamma$ point. 

\begin{figure}
\includegraphics[width=3.0in]{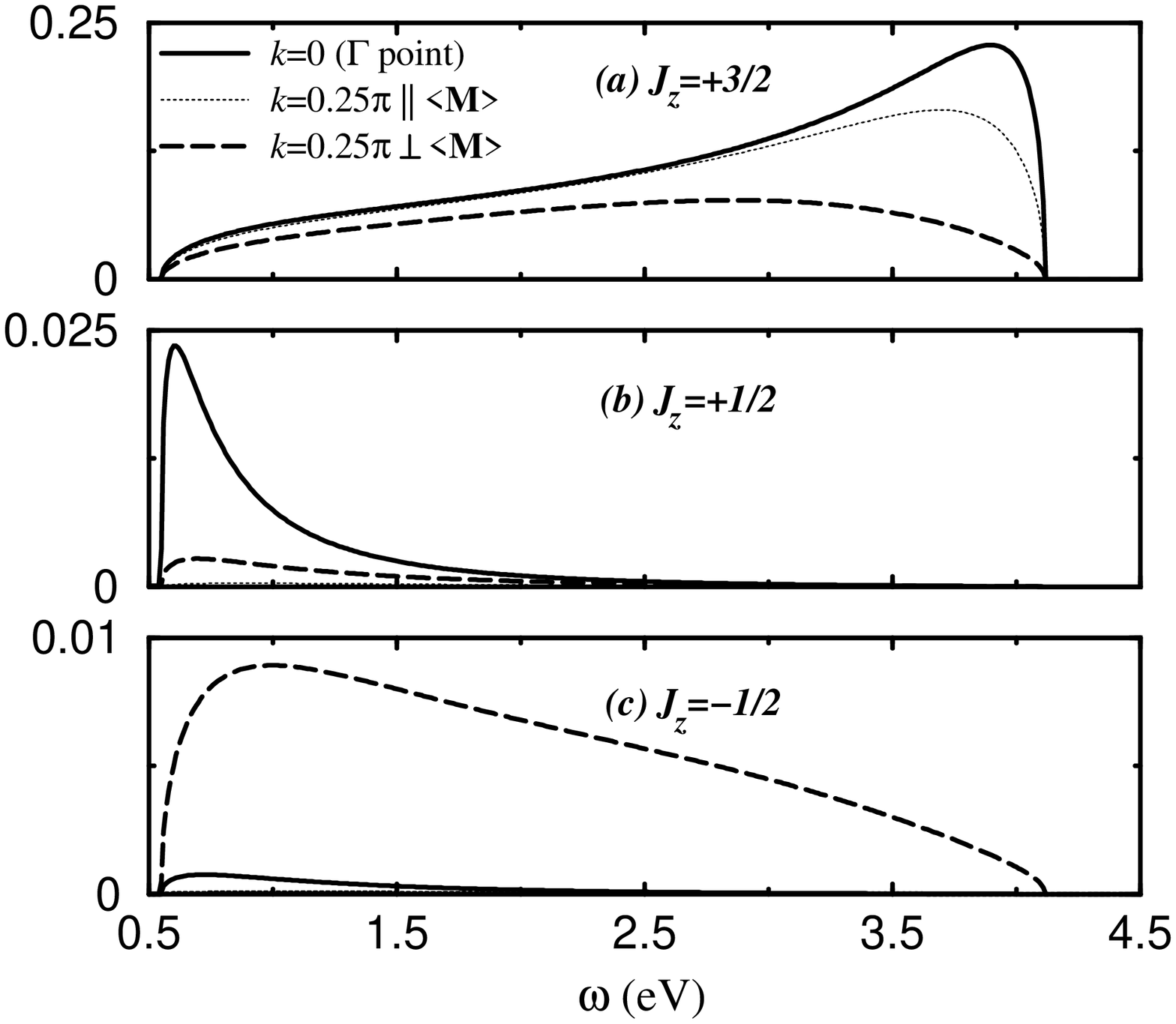}
\caption[a]{Variation of the $J_z$ components of the impurity band spectral function for 
$J_c=3.0\ eV$ and $T=0$ ((a) $J_z=+3/2$, (b) $J_z=+1/2$, and (c) $J_z=-1/2$) as 
the magnitude of the momentum {\it k} changes from 0 (solid curve) 
to ${\it k}=0.25\pi$ along the direction parallel (dotted curve) and perpendicular 
(long-dashed curve) to the average magnetization. Notice that the three 
graphs have different scales.}
\label{AnistrpImpBand}
\end{figure}
In the ferromagnetic phase, the quasi-particle lines split into four.
Since the finite magnetization competes with the chiral nature of the holes, 
the quasi-particle lines no longer correspond to well-defined chiral states.
This is confirmed by  Figs.~\ref{spectragraphs}(b)
and (c) where the curvature (related to the effective mass) 
of the quasi-particle bands depends on the direction of the momentum.
For the direction parallel to the average magnetization the spectrum of the 
quasiparticle band with $J_z=+1/2$ is shifted towards the semiconducting gap
by approximately $0.4 eV$ 
while the others bands are pushed towards negative energies. 
The positive shift in the $J_z=+1/2$ band is mainly due to the Zeeman splitting.
The band repulsion between this band and the impurity states is very small 
since the impurity band hardly includes $J_z=+1/2$ quasiparticles with momentum
parallel to the magnetization (see Fig.~\ref{AnistrpImpBand}). 
On the other hand the $J_z=+3/2$ quasiparticle band
shifts its spectrum by $\approx -0.2 eV$ due to 
strong repulsion with the $J_z=+3/2$ states at the impurity band.
However, for the direction perpendicular to the average magnetization 
the quasiparticle peak centered at $\omega  \approx 0.4 eV$ has 
${\bf J}\cdot{\hat{\bf k}}=+3/2$ character. 
Since the impurity band on the perpendicular plane
is formed mainly by $ {\bf J}\cdot{\hat{\bf k}}=\pm 1/2$ states, 
the  ${\bf J}\cdot{\hat{\bf k}}=+3/2$ quasiparticles do not suffer band repulsion.

Now examine the impurity band spectra in the insets.
In the paramagnetic phase the impurity band 
does not show significant variation with {\bf k}, indicating strong localization of the 
bound-state holes. In the ferromagnetic phase, the variation with {\bf k}, although small,
confirms that the bound-state holes are more mobile.
Typical fillings leading to the highest critical temperatures correspond to values of the 
chemical potential inside the impurity band. Therefore, the transport properties rely on the 
impurity band rather than on the quasi-particle bands.

As demonstrated by Fig.~\ref{AnistrpImpBand} for the ferromagnetic phase, the 
impurity-band spectral function along the direction parallel to the magnetization 
is predominantly composed of  $J_z=+3/2$ holes, whereas  
along the direction perpendicular to the magnetization, it is a mixture of 
$J_z=+3/2, +1/2,$ and $-1/2$ states.
This result suggests that the carrier polarization may be optimized 
by driving the current along the direction parallel to the 
magnetization. However, this may come with a price, since 
the carrier mobility is lower for the $J_z=+3/2$ heavy holes.
Due to the mixing of heavy and light holes,
the direction perpendicular to the magnetization may have higher mobility but
lower carrier-spin polarization. Also notice that the $J_z=-1/2$ states
participating in the local screening at the impurity band mostly display
finite perpendicular momentum. This is the configuration most energetically
favorable to avoid an exchange penalty  while fulfilling spin-orbit constraints.

In conclusion, we have calculated the spectra of the renormalized valence bands and the impurity 
band of Ga$_{1-x}$Mn$_{x}$As within the DMFA. We 
compare our results with existing ARPES data for the paramagnetic phase \cite{Okabayashi01}.
From the anisotropy of the impurity band in the ferromagnetic phase, 
we predict that the direction parallel to the magnetization will produce the most polarized 
spin current, whereas the perpendicular direction may display higher conductivity with lower 
polarization. It would be interesting to be able to compare our results in the ferromagnetic 
phase with additional ARPES data.

We acknowledge useful conversations with Paul Kent, Brian Moritz and Maciej Sawicki.
This research was supported 
by NSF grants DMR-0073308, DMR-0312680 and EPS-0132289 (ND EPSCoR), 
by the Department of Energy 
grant no. DE-FG03-03NA00071 (SSAAP program) and also
under contract DE-AC05-00OR22725 with Oak Ridge National Laboratory,
managed by UT-Battelle, LLC.
Research carried out in part at the Center for Functional
Nanomaterials, Brookhaven National Laboratory, which is supported by the
U.S. Department of Energy, Division of Materials Sciences and Division of
Chemical Sciences, under Contract No. DE-AC02-98CH10886.


\begin{thebibliography}{18}


\bibitem{Wolf01} S.\ A. Wolf, D.\ D. Awschalom, R.\ A. Buhrman, J.\ M. Daughton, 
S. von Moln\'ar, M.~L. Roukes, A. Y. Chtchelkanova, D. M. Treger, 
Science {\bf 294}, 1488 (2001).

\bibitem{Ohno99} Y. Ohno, D.~K. Young, B. Beschoten, F. Matsukura, H. Ohno and 
D.~D. Awschalom, Nature {\bf402}, 790 (1999).

\bibitem{MacDon05} For a recent review: A.~H. MacDonald, P. Schiffer, and N. Samarth, 
Nature Materials {\bf4}, 195 (2005).

\bibitem{Linnarsson97} M. Linnarsson, E. Janz\'en, B. Monemar, M. Kleverman, A. Thilderkvist, 
Phys. Rev. B {\bf 55}, 6938 (1997).

\bibitem{Okabayashi98} J. Okabayashi, A. Kimura, O. Rader, T. Mizokawa, and A. Fujimori, 
T. Hayashi and M. Tanaka, Phys. Rev. B {\bf 58}, R4211 (1998).

\bibitem{blakemore} J.\ S.\ Blakemore, J.\ Appl.\ Phys.\ {\bf 53}, R123 (1982). 

\bibitem{metzner-vollhardt} W.\ Metzner and D.\ Vollhardt, Phys.\ Rev.\ Lett.\
{\bf 62}, 324 (1989). 

\bibitem{muller-hartmann} E.~M\"uller-Hartmann, Z.\ Phys.\ {\bf B 74},
507--512 (1989).
 
\bibitem{pruschke} T.\ Pruschke, M.\ Jarrell and J.K.\ Freericks,
Adv.\ in Phys.\ {\bf 42}, 187 (1995).

\bibitem{georges} A.\ Georges, G.\ Kotliar,
W.\ Krauth and M.\ Rozenberg, Rev.\ Mod.\ Phys.\ {\bf 68}, 13 (1996).

\bibitem{Aryanpour04} K. Aryanpour, J. Moreno, M. Jarrell and R.~S. Fishman,
Phys. Rev. B {\bf 72}, 045343 (2005)

\bibitem{chattopadhyay} A.\ Chattopadhyay, S.\ Das Sarma and A.\ J.\ Millis, Phys.\ Rev.\ Lett.\ {\bf 87}, 227202 (2001).

\bibitem{Okabayashi01} J. Okabayashi, A. Kimura, O. Rader, T. Mizokawa, A. Fujimori, 
T. Hayashi and M. Tanaka, Phys. Rev. B {\bf 64}, 125304 (2001).

\bibitem{singley02} E.~J. Singley, R. Kawakami, D.~D. Awschalom, and D.~N. Basov, 
Phys. Rev. Lett. {\bf 89}, 097203 (2002).

\bibitem{singley03} E.~J. Singley, K.~S. Burch, R. Kawakami, J. Stephens, D.~D. Awschalom, 
and D.~N. Basov, Phys. Rev. B {\bf 68}, 165204 (2003).

\bibitem{burch05} K.~S. Burch,  E.~J. Singley, J. Stephens, R.~K. Kawakami, D.~D. Awschalom, 
and D.~N. Basov, Phys. Rev. B {\bf 71}, 125340 (2005).

\bibitem{burch04} K.~S. Burch, J. Stephens, R.~K. Kawakami, D.~D. Awschalom, and D.~N. Basov, 
Phys. Rev. B {\bf 70}, 205208 (2004).

\bibitem{Grandidier00} B. Grandidier, J.~P. Nys, C. Delerue, D. Sti\'evenard, Y. Higo, 
and M. Tanaka, Appl. Phys. Lett. {\bf 77}, 4001 (2000).
 
\bibitem{Tsuruoka02} T. Tsuruoka, N. Tachikawa, S. Ushioda, F. Matsukura, K. Takamura, and 
H. Ohno, Appl. Phys. Lett. {\bf 81}, 2800 (2002).

\bibitem{Sapega05} V.~F. Sapega, M. Moreno, M. Ramsteiner, L. Daweritz, and K.~H. Ploog, 
Phys. Rev. Lett. {\bf 94}, 137401 (2005).

\bibitem{janko} G.\ Zar\'and and B.\ Jank\'o, Phys.\ Rev.\ Lett. {\bf 89}, 047201 (2002).

\bibitem{Baldereschi73} A. Baldereschi and N.~O. Lipari,  Phys.\ Rev.\ B.\ {\bf 8}, 2697 (1973).

\bibitem{kacman} P.\ Kacman, Semicond.\ Sci. and Technol.\ {\bf 16}, R25 (2001).

\bibitem{taylor} D.\ W.\ Taylor, Phys.\ Rev.\ {\bf 156}, 1017 (1967); P.\ Soven, {\it ibid}. {\bf 156}, 809 (1967); 
P.\ L.\ Leath and B.\ Goodman, {\it ibid}. {\bf 148}, 968 (1966).

\bibitem{furukawa} N.\ Furukawa, J. Phys. Soc. Japan {\bf 63}, 3214 (1994) and 
Proc. Conference on Physics of Manganites (1998) (available at http://xxx.lanl.gov/abs/cond-mat/9812066).

\bibitem{Moreno05} J. Moreno, R.~S. Fishman and M. Jarrell, cond-mat/0507487.

\bibitem{mbe} H.\ Munekata, H.\ Ohno, S.\ von Moln\'ar, A.\ Segm\"uller, L.\ L.\ Chang and L.\ Esaki, Phys. Rev. Lett. {\bf 63}, 1849 (1989).

\bibitem{Ohno92} H.\ Ohno, H.\ Munekata, T.\ Penney, S.\ von Moln\'ar and L.\ L.\ Chang, Phys.\ Rev.\ Lett.\ {\bf 68}, 2664 (1992).

\bibitem{Ohno96} H.\ Ohno, A.\ Shen, F.\ Matsukura, A.\ Oiwa, A.\ Endo, S.\ Katsumoto and Y.\ Iye, Appl.\ Phys.\ Lett.\ {\bf 69}, 363 (1996).

\bibitem{VanEsch97} A.\ Van Esch, L.\ Van Bockstal, J.\ De Boeck, G.\ Verbanck, A.\ S.\ van Steenbergen, P.\ J.\ Wellmann, B.\ Grietens, R.\ Bogaerts, F.\ Herlach and G.\ Borghs, Phys.\ Rev.\ B.\ {\bf 56}, 13103 (1997).

\bibitem{Ohno98} H. Ohno, Science {\bf 281}, 951 (1998).

\bibitem{Chiba03} D. Chiba, K. Takamura, F. Matsukura and H. Ohno, 
Appl. Phys. Lett. {\bf 82}, 3020 (2003).


\bibitem{Edmonds04} K.~W. Edmonds, P. Boguslawski, K.~Y. Wang, R.~P. Campion,
S.~N. Novikov, N.~R.~S. Farley, B.~L. Gallagher, C.~T. Foxon, M. Sawicki,
T. Dietl, M. Buongiorno Nardelli, and J. Benholc, Phys. Rev. Lett. {\bf 92},
037201 (2004).



\end{thebibliography}
\end{document}